\documentclass{article}

\usepackage{hyperref}
\usepackage{color, soul}
\usepackage{fullpage}
\usepackage{amsfonts}
\usepackage{amssymb}
\usepackage{amsmath}
\usepackage{latexsym}
\usepackage{amsthm}
\usepackage{eepic}
\usepackage{textcomp}
\usepackage{stmaryrd}

\theoremstyle{plain}
\newtheorem{theorem}{Theorem}[section]
\newtheorem{lemma}[theorem]{Lemma}

\newtheorem{definition}[theorem]{Definition}

\newtheorem{fact}[theorem]{Fact}

\newcommand{\microspace}{\mspace{0.5mu}}

\newcommand{\defeq}{\stackrel{\mathrm{def}}{=}}
\newcommand{\tr}{\operatorname{Tr}}

\def\real{\mathbb{R}}

\def \lket {\left|}
\def \rket {\right\rangle}
\def \lbra {\left\langle}
\def \rbra {\right|}
\newcommand{\ket}[1]{\lket\microspace #1 \microspace\rket}
\newcommand{\bra}[1]{\lbra\microspace #1 \microspace\rbra}
\newcommand{\ketbra}[1]{\lket\microspace #1 \microspace\rket\lbra\microspace #1 \microspace\rbra}

\def\X{\mathcal{X}}
\def\Y{\mathcal{Y}}

\def\A{\mathcal{A}}

\def\D{\mathcal{D}}
\def\E{\mathcal{E}}

\def\L{\mathcal{L}}

\def\H{\mathcal{H}}

\def\T{\mathcal{T}}

\def\c{\mathsf{cap}}
\def\tc{\widetilde{\mathsf{cap}}}
\def\id{\mathbb{I}}

\newcommand{\suppress}[1]{}

\newcommand {\br} [1] {\ensuremath{ \left( #1 \right) }}
\newcommand {\Br} [1] {\ensuremath{ \left[ #1 \right] }}

\newcommand {\minusspace} {\: \! \!}
\newcommand {\smallspace} {\: \!}
\newcommand {\fn} [2] {\ensuremath{ #1 \minusspace \br{ #2 } }}
\newcommand {\Fn} [2] {\ensuremath{ #1 \minusspace \Br{ #2 } }}

\newcommand {\mutinf} [2] {\fn{\mathrm{I}}{#1 \smallspace : \smallspace #2}}
\newcommand {\condmutinf} [3] {\mutinf{#1}{#2 \smallspace \middle\vert \smallspace #3}}

\newcommand {\relent} [2] {\fn{\mathrm{S}}{#1 \middle\| #2}}
\newcommand {\ent} [1] {\fn{\mathrm{S}}{#1}}

\DeclareMathOperator*{\bigE}{\mathbb{E}}

\newcommand {\expec} [2] {\Fn{\bigE_{\substack{#1}}}{#2}}

\title{Chain-rules for channel capacity}
\author{
Rahul Jain
\thanks{
Centre for Quantum Technologies and Department of Computer Science, National University of Singapore and MajuLab, UMI 3654, Singapore. Email:~\tt{rahul@comp.nus.edu.sg}}
}
\date{}

\begin{document}
\maketitle
\begin{abstract}
We show some chain-rules for the  capacity\footnote{In some sense, the maximum amount of information that can be conveyed through the channel.} of classical-quantum and quantum channels. We use the concept of  Nash-Equilibrium in  game-theory, and its existence in suitably defined games, to arrive at the chain-rules.
\end{abstract}

\section{Introduction}
 Given a quantum state $\rho_{AB}$, the mutual-information between the registers $A$ and $B$ is defined as $$\mutinf{A}{B}_\rho \defeq \ent{\rho_A} + \ent{\rho_B} - \ent{\rho_{AB}}\enspace,$$ where
$\ent{\cdot}$ represents the von Neumann entropy.  An important property satisfied by mutual-information, often referred as {\em the chain-rule} is that given state $\rho_{ABC}$,
$$ \mutinf{AB}{C}_\rho =  \mutinf{A}{C}_\rho  + \condmutinf{B}{C}{A}_\rho\enspace.$$
This property forms a basis for information-theoretic methods and has been crucially used in showing several {\em direct-sum} and {\em direct-product} results in {\em communication complexity};  a partial list includes~\cite{Razborov1992,Parnafes1997,Nayak99,ChakrabartiSWY01,BJKS02,Shaltiel2003,JainRS03,JainRS03a,JainRS05,Klauck2007,Lee2008,Viola2008,JainKN08,HarshaJMR10,Klauck2010,Jain15,Ben-Aroya2008,JainY12,Sherstov2012,BravermanW15,BravermanR14,BarakBCR13,BravermanRWY13,Braverman2013a,Braverman15,JainPY16,Gillat16,BravermanG18,Sherstov18}.

\subsection*{Our results} Let $\H_M$ be a Hilbert space and $g: \X \rightarrow \D(M)$ be a {\em classical-quantum} (c-q) channel, where $\X$ is a finite set and $\D(M)$ is the set of quantum states supported on $\H_M$.  The capacity of $g$ is defined as 
$$\c(g) \defeq \max_{\mu \in \D(\X)} \expec{x \leftarrow \mu }{\relent{g(x)}{g_\mu}}\enspace,$$
where $\relent{\cdot}{\cdot}$ represents the relative-entropy, $\D(\X)$ is the set of probability distributions supported on $\X$ and $\rho_\mu = \expec{x\leftarrow\mu}{\rho(x)}$.

We ask if capacity can be shown to satisfy some chain-rules and answer in the affirmative. Let  $g : \X \rightarrow  \D(M)$ be a c-q channel where $\X = (\X_1 \times \ldots \times  \X_k)$. For $i\in[k]$ and $\mu \in \D(\X)$, define channel $g^i_{\mu}: \X_i \rightarrow \D(M)$ given by $g^i_{\mu}(x_i)=\expec{x'\leftarrow \mu}{g(x_i,x'_{-i})}$.  We show the following chain-rule.
\begin{theorem}[Chain-rule-1 for c-q capacity] \label{thm:chain1}
$$\c(g) \ge \sum_{i=1}^k \min_{ \theta \in \D(\X)} \c(g_\theta^i) \enspace. $$
\end{theorem} 

Next we show a stronger chain-rule using product-distributions and with a change of quantifiers.  Let $\A=\D(\X_1) \times \ldots \times \D(\X_k)$.
\begin{theorem}[Chain-rule-2 for c-q capacity] \label{thm:chain2}
\begin{align*}
\c(g) \ge \min_{\theta \in \A}\sum_{i=1}^k \c(g^i_\theta)\enspace. 
\end{align*}
\end{theorem}
We generalize the above to allow for conditioning on some events in Theorem~\ref{thm:chain3}.

Next we consider a chain-rule for quantum channels. Let $\H_A, \H_B, \H_C$ be Hilbert spaces such that $\H_A$ is isomorphic to  $\H_B$. Let $g: \L(B)\rightarrow \L(C)$ be a quantum channel, where $\L(B)$ is the set of all linear operators supported on $\H_B$. For $\rho \in \D(A)$, let $\hat{\rho}\in \D(AB)$ represent the {\em canonical purification} of $\rho$. The capacity of $g$ is defined as 
$$\c(g) \defeq \max_{\rho \in \D(A)} \mutinf{A}{C}_{g(\hat{\rho})} \enspace.$$

Let $\H_A = \H_{A_1} \otimes \ldots \otimes \H_{A_k}, \H_B = \H_{B_1} \otimes \ldots \otimes \H_{B_k}$ and $\H_C$ be Hilbert spaces such that forall $i\in [k]$, $\H_{A_i}$ is isomorphic to $\H_{B_i}$. Let $g: \L(B)\rightarrow \L(C)$ be a quantum channel.  Let $\A=\D(A_1) \times \ldots \times \D(A_k)$. We use $(\rho_1,  \ldots, \rho_k)  \in \A$ to represent $(\rho_1 \otimes \ldots \otimes  \rho_k)$. For $i\in[k]$, let $\hat{\rho_i}\in \D(A_iB_i)$ represent the canonical purification of $\rho_i$. For $i\in[k]$ and $\theta \in \D(A)$, define channel $g^i_{\theta}: \L(B_i) \rightarrow \L(C)$ given by $g^i_\theta(\rho_i)=g(\rho_i,\theta_{-i})$. We show the following chain-rule.
\begin{theorem}[Chain-rule for quantum capacity] \label{thm:chain4}
\begin{align*}
\c(g) ~\ge~   \min_{\theta \in \A}\sum_{i=1}^k \c(g^i_\theta)\enspace.
\end{align*}
\end{theorem}
We use the concept of  {\em Nash-Equilibrium} in game-theory, and its existence in suitably defined games, to show Theorems~\ref{thm:chain2},~\ref{thm:chain3} and~\ref{thm:chain4}.
\subsection*{Applications} Theorem~\ref{thm:chain1} has been used by Jain and Klauck~\cite{JainK09} to show  direct-sum results for classical and quantum {\em Simultaneous-Message-Passing} (SMP) models of communication complexity, with no shared resource between parties with inputs. We hope that these chain-rules are able to find other interesting applications.
\subsection*{Organization} In Section~\ref{sec:Preliminaries} we present some information theoretic preliminaries. In Section~\ref{sec:chianrule} we present the chain-rules for the capacity of c-q channels. In Section~\ref{sec:chainquantum}, we present the chain-rule for the capacity of quantum channels.

\section{Preliminaries}
\label{sec:Preliminaries}

\subsection*{Information theory}
All logarithms are evaluated to the base $2$. For a finite set $\X$, let $\D(\X)$ be the set of all probability distributions supported on $\X$.  For  $\mu\in\D(\X)$, let $\mu(x)$ represent the probability of $x\in\X$ according to
$\mu$. We use the same symbol to represent a random variable $X$ and its distribution whenever it is clear from the context. 

Consider a finite dimensional Hilbert space $\H$ endowed with an inner product $\langle \cdot, \cdot \rangle$ (we use the standard {\em ket-bra} notation). A quantum state (or a density matrix or a state) is a positive semi-definite matrix supported on $\H$ with trace equal to $1$. It is called pure if and only if its rank is $1$. Let $\ket{\psi}$ be a unit vector on $\H$, that is $\langle \psi,\psi \rangle=1$.  With some abuse of notation, we use $\psi$ to represent the state and also the density matrix $\ketbra{\psi}$, associated with $\ket{\psi}$. 
 
A quantum register $A$ is associated with some Hilbert space $\H_A$. 
We denote by $\mathcal{D}(A)$, the set of quantum states on the Hilbert space $\H_A$. State $\rho$ with subscript $A$ indicates $\rho_A \in \mathcal{D}(A)$. If two registers $A,B$ are associated with isomorphic Hilbert spaces, we represent the relation by $A\equiv B$.  Composition of two registers $A$ and $B$, denoted $AB$, is associated with Hilbert space $\H_A \otimes \H_B$.  For two quantum states $\rho\in \mathcal{D}(A)$ and $\sigma\in \mathcal{D}(B)$, $\rho\otimes\sigma \in \mathcal{D}(AB)$ represents the tensor product (Kronecker product) of $\rho$ and $\sigma$. The identity operator on $\H_A$ is denoted $\id_A$. Let $\rho_{AB} \in \mathcal{D}(AB)$. We define
\[ \rho_{B} \defeq \tr_{A}{\rho_{AB}}
\defeq \sum_i (\bra{i} \otimes \id_{B})
\rho_{AB} (\ket{i} \otimes \id_{B})\enspace, \]
where $\{\ket{i}\}_i$ is an orthonormal basis for $\H_A$. The state $\rho_B\in \mathcal{D}(B)$ is referred to as the marginal state of $\rho_{AB}$. Unless otherwise stated, a missing register from subscript in a state will represent partial trace over that register. Given a $\rho_A\in\mathcal{D}(A)$, a  purification of $\rho_A$ is a pure state $\rho_{AB}\in \mathcal{D}(AB)$ such that $\tr_{B}{\rho_{AB}}=\rho_A$. Purification of a quantum state is not unique. Suppose $A\equiv B$. Given $\{\ket{i}_A\}$ and $\{\ket{i}_B\}$ as orthonormal bases over $\H_A$ and $\H_B$ respectively, the canonical purification of a quantum state $\rho_A$ is $(\rho_A^{\frac{1}{2}}\otimes\id_B)\left(\sum_i\ket{i}_A\otimes\ket{i}_B\right)$. 

\begin{definition} \label{def:inf}
\begin{enumerate}
\item For a natural number  $k$, let $[k]\defeq\{1, \ldots,
k\}$.  For $i\in[k]$ let $-i\defeq[k]-\{i\}; <i\defeq[i-1]$. For string $x=(x_1, \ldots, x_k)$ and $T\subseteq[k]$, let $x_T$ be sub-string of $x$ with indices in $T$. For string $x=(x_1, \ldots, x_k)$ and $i\in[k]$, define $(x_i,x_{-i})\defeq x$. Similarly for tuples of random variables and quantum states. 
\item The expectation of function $f$  according to distribution $\mu$ is defined
as $$\expec{x \leftarrow \mu}{f(x)} \defeq\sum_{x\in\X} \mu(x)
\cdot f(x)\enspace. $$ 
\item  For $\mu, \lambda \in \D(\X)$, the distribution $\mu
\otimes \lambda$ is defined as
$(\mu\otimes\lambda)(x_1,x_2)\defeq\mu(x_1)\cdot\lambda(x_2)$.  We use $(\mu, \lambda)$ to represent $\mu \otimes \lambda$. Let $\mu^k$ represent the $k$-tuple $(\mu, \cdots, \mu)$.
\item For jointly distributed random variables $XY$ distributed according to $\mu$, denoted $XY\sim \mu$, define $Y_x\defeq(Y|X=x)$ and let $\mu_x$ represent the distribution of $Y_x$. 
\item The entropy of a state $\rho$ is given by: $\ent{\rho} = -\tr(\rho\log \rho).$
\item 
The relative-entropy between two states $\rho$ and $\sigma$ is given by: $\relent{\rho}{\sigma} = \tr(\rho\log\rho) - \tr(\rho\log\sigma).$
\item A state of the form
\[ \rho_{XY} = \sum_x p(x)\ketbra{x}_X\otimes\rho^x_{Y}\]
is called a classical-quantum (c-q) state, with $X$ being a classical register/random variable and $Y$ being a quantum register.
\item The mutual-information between $Y$ and $Z$ with respect to a state $\rho_{YZ}$ is defined as
\[  \mutinf{Y}{Z}_\rho = \relent{\rho_{YZ}}{\rho_Y\otimes\rho_Z}.\]
\item The conditional-mutual-information between $Y$ and $Z$ conditioned on $X$ with respect to a state $\rho_{XYZ}$, is defined as
\[  \condmutinf{Y}{Z}{X}_\rho= \mutinf{XY}{Z}_\rho - \mutinf{X}{Z}_\rho.\]
\item A quantum channel $g: \mathcal{L}(A)\rightarrow \mathcal{L}(B)$ is a completely positive and trace preserving (CPTP) linear map. 
\item Let $\H_M$ be a Hilbert space, $g: \X \times  \Y \rightarrow \D(M)$ be a
classical-quantum (c-q) channel and $\mu \in \D(\X \times \Y)$. Define 
$$g_\mu(x) = \expec{y \leftarrow \mu_x}{g(x,y)} ~;~  g_\mu(y) = \expec{x \leftarrow \mu_y}{g(x,y)} ~;~ g_\mu = \expec{(x,y) \leftarrow \mu}{g(x,y)} \enspace.$$
\end{enumerate}
\end{definition}
We need the following facts. 
\begin{fact}[Chain-rule for mutual-information]\label{fact:chaininf} For a state $\rho_{X_1 \ldots X_k M}$:
$$\mutinf{X_1 \ldots X_k}{M}_\rho=\sum_{i=1}^k\condmutinf{X_i}{M}{X_{<i}}_\rho\enspace.$$
If $\rho_{X_1, \ldots, X_k} = \rho_{X_1} \otimes \ldots \otimes \rho_{X_k}$, then:~
$$\mutinf{X_1 \ldots X_k}{M}_\rho \ge\sum_{i=1}^k\mutinf{X_i}{M}_\rho\enspace.$$
\end{fact}

\begin{fact}[Joint-convexity for relative-entropy]\label{fact:rel-conv}
For states $\rho, \rho', \sigma, \sigma'$ and $p\in[0,1]$,
$$ \relent{p \rho + (1-p) \rho'}{p\sigma + (1-p) \sigma'} \leq p \cdot
\relent{\rho}{\sigma} + (1-p) \cdot \relent{\rho'}{\sigma'} \enspace.$$
\end{fact}
\suppress{
\begin{fact}[Chain-rule for relative-entropy]
    \label{fact:relchain}
    \begin{align*}
       \relent{XY}{X'Y'} &= \relent{X}{X'}
    + \expec{x\leftarrow X}{\relent{Y_x}{Y'_x}} \enspace.
    \end{align*} 
    In particular, using Fact~\ref{fact:rel-conv}:
    $$ \relent{XY}{X'\otimes Y'}
    = \relent{X}{X'} + \expec{x\leftarrow X}{\relent{Y_x}{Y'}}
    \geq \relent{X'}{X} + \relent{Y'}{Y}\enspace.$$
\end{fact}
}
\begin{fact}[see e.g Fact 2.5~\cite{JainPY16}]\label{fact:inf} For state $\rho_{XY}$: 
\begin{align*}
    \mutinf{X}{Y}_\rho
    =\relent{\rho_{XY}}{\rho_X \otimes \rho_Y} = \min_{\sigma_Y}\relent{\rho_{XY}}{\rho_X \otimes \sigma_Y} = \min_{\sigma_X,\sigma_Y} \relent{\rho_{XY}}{\sigma_X \otimes \sigma_Y}  \enspace.
\end{align*}
\end{fact}
\begin{fact}[Chain-rule for relative-entropy]\label{fact:relchain}
For c-q states $\rho_{XY}, \sigma_{XY}$ ($X$ being classical):
$$ \relent{\rho_{XY}}{\sigma_{XY}} = \relent{\rho_X}{\sigma_X} + \expec{x \leftarrow X}{\relent{\rho_Y^x}{\sigma^x_Y}}\enspace.$$
\end{fact}
\begin{fact}[Pretty-Good-Measurement] \label{fact:PGM}
Let $p\in[0,1]$ and 
$$\rho_{XA} =  p\cdot \ketbra{0}_X \otimes \rho^0_A + (1-p)\cdot \ketbra{1}_X \otimes \rho^1_A \enspace.$$ 
There exists a quantum map $\E$ such that $\E(\rho_A)=\rho_{XA}$.
\end{fact}
\begin{fact}[Monotonicity]\label{fact:mono}
Let $\rho, \sigma$ be states and $\E$ be a quantum map. Then,
$$\relent{\rho}{\sigma} \ge  \relent{\E(\rho)}{\E(\sigma)} \enspace.$$
Let $\rho_{AB}$ be a state and $\E:\L(B) \rightarrow \L(C)$ be a quantum map. Then,
$$\mutinf{A}{B}_\rho \ge \mutinf{A}{C}_{\E(\rho)}\enspace.$$
\end{fact}
\subsection*{Game theory}
We use the following powerful  theorem from game theory, which
is a consequence of the {\em Kakutani fixed-point theorem} in real analysis.
\begin{fact} [Nash-Equilibrium, Proposition~20.3~\cite{OsborneR94}]
\label{fact:nash}
Let $k, n$ be a positive integers. Let $\A = \A_1 \times \ldots \times \A_k$, where each $\A_i$ is a non-empty, convex and compact subset of $\real^n$. For each $i \in [k]$, let $u_i: \A\rightarrow \real$ be a continuous function such that
\begin{displaymath}
\forall a= (a_1,  \ldots, a_k) \in \A \;:\; \mbox{the set } 
\{a'_i\in\A_i \;:\;                u_i(a'_i,a_{-i})  \ge u_i(a)\} \mbox{ is convex}.
\end{displaymath}
There is an {\em equilibrium point} $a^\ast \in \A$ such that 
\begin{displaymath}
\forall i\;:\; \max_{a_i\in \A_i}\, u_i(a_i,a_{-i}^\ast) 
\quad = \quad u_i(a^\ast)\enspace.
\end{displaymath}
\end{fact}
Following {\em min-max} theorem is a corollary of the above. 
\begin{fact} [Min-Max, Proposition~22.2~\cite{OsborneR94}]
\label{fact:minmax}
Let $k, n$ be a positive integers. Let $\A = \A_1 \times \A_2$, where each $\A_i$ is a non-empty, convex and compact subset of $\real^n$. Let $u: \A\rightarrow \real$ be a continuous function such that $\forall a= (a_1, a_2) \in \A$, the sets, 
\begin{align*}
\{a'_1\in\A_1 \;:\;                u(a'_1,a_{2})  \ge u(a)\} \mbox{ and }
\{a'_2\in\A_2 \;:\;                u(a_1,a'_{2})  \le u(a)\} \mbox{ are convex}\enspace.
\end{align*}
There is a min-max point $(a_1^\ast,a_2^\ast) \in \A$ such that 
\begin{displaymath}
\min_{a_2\in \A_2}\, \max_{a_1\in \A_1}\,  u(a_1,a_{2}) 
\quad = \quad \max_{a_1\in \A_1}\, \min_{a_2\in \A_2}\, u(a_1,a_{2}) 
\quad = \quad u(a_1^\ast,a_2^\ast)\enspace.
\end{displaymath}
\end{fact}
\section{Chain rules for the capacity of c-q channels} \label{sec:chianrule}
\subsection*{Capacity}
Let $\H_M$ be a Hilbert space and $g: \X \rightarrow \D(M)$ be a c-q channel. 
\begin{definition}[Capacity]
\label{def:cap}
The capacity of $g$ is defined as 
$$\c(g) \defeq \max_{\mu \in \D(\X)} \expec{x \leftarrow \mu }{\relent{g(x)}{g_\mu}}\enspace.$$
\end{definition}
Jain~\cite{Jain06} considered the following notion of a {\em capacity-dual}. 
\begin{definition}[Capacity-dual]
The capacity-dual of $g$ is defined as 
$$\tc(g) \defeq \min_{\gamma \in \D(\X)}\max_{x \in \X} ~\relent{g(x)}{g_\gamma}\enspace.$$
\end{definition}
Using Fact~\ref{fact:rel-conv} and Fact~\ref{fact:minmax}, Jain~\cite{Jain06} showed  that capacity is lower bounded by capacity-dual.
\begin{fact}[Lemma 2.~\cite{Jain06}] \label{fact:capdual}
$$\c(g) \ge \max_{\mu \in \D(\X)} \min_{\gamma \in \D(\X)} \expec{x \leftarrow \mu}{\relent{g(x)}{g_\gamma}} = \min_{\gamma \in \D(\X)}\max_{x \in \X} \relent{g(x)}{g_\gamma} = \tc(g)\enspace.$$
\end{fact}
We show that they are in fact the same.
\begin{lemma}\label{lem:capsamedual}
$\c(g)=\tc(g)$.
\end{lemma}
\begin{proof}
Consider,
\begin{align*}
    \c(g) &= \max_{\mu \in \D(\X)} \expec{x \leftarrow \mu}{\relent{g(x)}{g_\mu}} \\
    &\le  \max_{\mu \in \D(\X)} \min_{\sigma\in\D(M)} \expec{x \leftarrow \mu}{\relent{g(x)}{\sigma}}&\mbox{(Facts~\ref{fact:inf},~\ref{fact:relchain})}\\
    &\le\tc(g)\enspace.
\end{align*}
Combined with Fact~\ref{fact:capdual} shows the desired.
\end{proof}
\subsection*{Chain-rules}
 Let  $g : \X \rightarrow  \D(M)$ be a channel where $\X = (\X_1 \times \ldots \times  \X_k)$. For $i\in[k]$ and $\mu \in \D(\X)$, define channel $g^i_{\mu}: \X_i \rightarrow \D(M)$ given by $g^i_{\mu}(x_i)=g_{\mu}(x_i)$. We start with the following chain-rule.
\begin{theorem}[Chain-rule-1 for c-q capacity] 
$$\c(g) \ge \sum_{i=1}^k \min_{ \theta \in \D(\X)} \c(g_\theta^i) \enspace. $$
\end{theorem} 
\begin{proof}
We show the result for $k=2$.  The result for larger $k$ follows by induction. Let's rename $\X_1$ as $\X$ and $\X_2$ as $\Y$. For $x \in \X$, let $g^x : \Y \rightarrow \D(M)$ be a channel given by $g^x(y) = g(x,y)$. Let ${\mu_x}$ be a
distribution on $\Y$ such that $\c(g^x) = \expec{y \leftarrow \mu_x }{\relent{g^x(y)}{g^x_{\mu_x}}}$. Let $g^X : \X \rightarrow \D(M)$ be a channel given by $g^X(x) = \expec{y \leftarrow \mu_x}{g(x,y)}$. Let ${\mu_X}$ be a distribution on $\X$ such that $\c(g^X) = \expec{x \leftarrow \mu_X }{\relent{g^X(x)}{g^X_{\mu_X}}}$. Let $\rho_{XYM}$ be a c-q state with $XY$ classical, such that $\forall (x,y):~ X \sim \mu_X;~ (Y~|~X=x) \sim \mu_x;~ \rho_M^{xy}= g(x,y)$.
From Fact~\ref{fact:inf}
\begin{align*}
    \c(g) &\ge \mutinf{XY}{M}_\rho &\mbox{(Definition~\ref{def:cap})}\\
    &= \mutinf{X}{M}_\rho + \condmutinf{Y}{M}{X}_\rho &\mbox{(Fact~\ref{fact:chaininf})}\\
&= \c(g^X) + \expec{x\leftarrow X}{\c(g^x)}\\
&\ge \min_{ \theta \in \D(\X \times \Y)} \c(g_\theta^1) + \min_{ \theta \in \D(\X \times \Y)} \c(g_\theta^2) \enspace. 
\end{align*}
\end{proof}
Next we show a stronger chain-rule using product-distributions and with a change of quantifiers. Let $\A=\D(\X_1) \times \ldots \times \D(\X_k)$.
\begin{theorem}[Chain-rule-2 for c-q capacity] 
\begin{align*}
\c(g) &\ge  \min_{(\theta,\gamma) \in \A \times \A} \sum_{i=1}^k \max_{x_i}  { \relent{g_{\theta}(x_i)}{g_{\theta_{-i},\gamma_i}}}  \\
&=  \min_{\theta \in \A}\sum_{i=1}^k \c(g^i_\theta)\enspace. & \mbox{(Lemma~\ref{lem:capsamedual})}
\end{align*}
\end{theorem}
\begin{proof}
For all $i\in[k] , \mu= (\mu_1, \ldots , \mu_k) \in \A$, define 
\begin{displaymath}
    u_i(\mu) = \min_{\gamma_i \in \D(\X_i)} \expec{x_i \leftarrow\mu_i}{ \relent{g_{ \mu}(x_i)}{g_{\mu_{-i},\gamma_i}}}\enspace. 
\end{displaymath}
For all $\mu, \mu'_i, \mu''_i, p\in [0,1]$,
\begin{align}
& u_i(p\mu'_i + (1-p)\mu_i'', \mu_{-i}) \nonumber\\
& =   \min_{\gamma_i}  \expec{x_i \leftarrow p\mu'_i + (1-p)\mu_i''}{ \relent{g_{\mu}(x_i)}{g_{\mu_{-i},\gamma_i}}} \nonumber \\
& = \min_{\gamma_i}  \left( p \expec{x_i \leftarrow \mu'_i}{ \relent{g_{\mu}(x_i)}{g_{\mu_{-i},\gamma_i}}} + (1-p) \expec{x_i \leftarrow \mu_i''}{ \relent{g_{\mu}(x_i)}{g_{\mu_{-i},\gamma_i}}} \right)\nonumber \\
& \ge p \left(\min_{\gamma_i}   \expec{x_i \leftarrow \mu'_i}{ \relent{g_{\mu}(x_i)}{g_{\mu_{-i},\gamma_i}}}\right) + (1-p)  \left(\min_{\gamma_i}\expec{x_i \leftarrow \mu_i''}{\relent{g_{\mu}(x_i)}{g_{\mu_{-i},\gamma_i}}}\right) \nonumber \\
& = p \cdot u_i(\mu'_i, \mu_{-i}) + (1-p) \cdot u_i(\mu_i'',\mu_{-i}) \enspace. \label{eq:nashcond}
\end{align}
From Eq.~\eqref{eq:nashcond} and Fact~\ref{fact:nash} (by letting $\forall i: (\A_i, u_i) \leftarrow (\D(\X_i), u_i)$), we get $\theta = (\theta_1, \ldots, \theta_k) \in \A$ such that, 
\begin{align*}
\forall i:~u_i(\theta) &= \max_{\mu_i\in \D(\X_i)} u_i(\mu_i,\theta_{-i}) \\
&= \max_{\mu_i} \min_{\gamma_i} \expec{x_i \leftarrow\mu_i}{ \relent{g_{\theta}(x_i)}{g_{\theta_{-i},\gamma_i}}} \\
&= \min_{\gamma_i} \max_{x_i}  { \relent{g_{\theta}(x_i)}{g_{\theta_{-i},\gamma_i}}}\enspace.
 & \mbox{(Fact~\ref{fact:capdual})}
\end{align*}
Let $\rho_{XM}$ be a c-q state with $X$ classical, such that $X=(X_1 \ldots X_k) \sim \theta$ and $\forall x\in \X:~  \rho_M^{x}= g(x)$. Consider,
\begin{align*}
     \sum_{i=1}^k\min_{\gamma_i} \max_{x_i}  { \relent{g_{\theta}(x_i)}{g_{\theta_{-i},\gamma_i}}} &= \sum_i u_i(\theta) \\
     &= \sum_i \min_{\gamma_i} \expec{x_i \leftarrow \theta_i}{ \relent{g_{\theta}(x_i)}{g_{\theta_{-i},\gamma_i}}}\\
    &\le \sum_i \expec{x_i \leftarrow \theta_i}{ \relent{g_{\theta}(x_i)}{g_{\theta_{-i},\theta_i}}} \\
    &= \sum_i \mutinf{X_i}{M}_\rho &\mbox{(Fact~\ref{fact:inf})}\\
    & \le \mutinf{X}{M}_\rho &\mbox{(Fact~\ref{fact:chaininf})} \\
    & \le \c(g) \enspace. & \mbox{(Definition~\ref{def:cap})} 
\end{align*}
This concludes the desired. 
\end{proof}
We generalize the above to allow for conditioning on some events. Let 
$$\T=\{(T,x_T)~|~ T \subseteq [k], x_T \in \X_T\}.$$ Below whenever $i \in T$, define $\relent{\cdot}{\cdot} \defeq 0$.
\begin{theorem}[Chain-rule-3 for c-q capacity] \label{thm:chain3}
\begin{align*}
\c(g) &\ge \max_{\alpha \in \D(\T)} \min_{(\theta,\gamma) \in \A \times \A} \sum_{i=1}^k \max_{x_i}  \expec{(T,x_T) \leftarrow \alpha}{ \relent{g_{ \theta}(x_i, x_T)}{g_{\theta_{-i},\gamma_i}(x_T)}}\enspace.
\end{align*}
\end{theorem}
\begin{proof}
Let $\alpha\in\D(\T)$. For all $i\in[k] , \mu= (\mu_1, \ldots, \mu_k) \in \A$, define,
\begin{displaymath}
    u_i(\mu) = \min_{\gamma_i \in \D(\X_i)} \expec{(T,x_T) \leftarrow \alpha, x_i \leftarrow\mu_i}{ \relent{g_{\mu}(x_i, x_T)}{g_{\mu_{-i},\gamma_i}(x_T)}} \enspace. 
\end{displaymath}
For all $\mu, \mu'_i, \mu''_i, p\in [0,1]$,
\begin{align}
u_i(p\mu'_i + (1-p)\mu_i'', \mu_{-i}) & =   \min_{\gamma_i}  \expec{(T,x_T) \leftarrow \alpha, x_i \leftarrow p\mu'_i + (1-p)\mu_i''}{ \relent{g_{\mu}(x_i,x_T)}{g_{\mu_{-i},\gamma_i}(x_T)}} \nonumber\\
& = \min_{\gamma_i} \biggl( p \expec{(T,x_T) \leftarrow \alpha, x_i \leftarrow \mu'_i}{ \relent{g_{\mu}(x_i,x_T)}{g_{\mu_{-i},\gamma_i}(x_T)}} \nonumber \\
& \qquad + (1-p) \expec{(T,x_T) \leftarrow \alpha, x_i \leftarrow \mu_i''}{ \relent{g_{\mu}(x_i,x_T)}{g_{\mu_{-i},\gamma_i}(x_T)}} \biggr)\nonumber\\
& \ge p \left(\min_{\gamma_i}   \expec{(T,x_T) \leftarrow \alpha, x_i \leftarrow \mu'_i}{ \relent{g_{\mu}(x_i,x_T)}{g_{\mu_{-i},\gamma_i}(x_T)}}\right) \nonumber\\
& \qquad + (1-p) \left(\min_{\gamma_i}\expec{(T,x_T) \leftarrow \alpha, x_i \leftarrow \mu_i''}{ \relent{g_{\mu}(x_i,x_T)}{g_{\mu_{-i},\gamma_i}(x_T)}}  \right) \nonumber\\
& = p \cdot u_i(\mu'_i, \mu_{-i}) + (1-p) \cdot u_i(\mu_i'', \mu_{-i}) \enspace.\label{eq:nashcond3}
\end{align}
From Eq.~\eqref{eq:nashcond3} and Fact~\ref{fact:nash} (by letting $\forall i: (\A_i, u_i) \leftarrow (\D(\X_i), u_i)$), we get $\theta = (\theta_1, \ldots, \theta_k) \in \A$ such that, 
\begin{align}
\forall i:~u_i(\theta) &= \max_{\mu_i\in \D(\X_i)} u_i(\mu_i,\theta_{-i}) \nonumber\\
&= \max_{\mu_i} \min_{\gamma_i} \expec{(T,x_T) \leftarrow \alpha,  x_i \leftarrow\mu_i}{ \relent{g_{\theta}(x_i, x_T)}{g_{\theta_{-i},\gamma_i}(x_T)}} \nonumber\\
&= \min_{\gamma_i } \max_{x_i}  \expec{(T,x_T) \leftarrow \alpha}{ \relent{g_{ \theta}(x_i, x_T)}{g_{\theta_{-i},\gamma_i}(x_T)}}\enspace.& \mbox{(Fact~\ref{fact:rel-conv} and Fact~\ref{fact:minmax})} \label{eq:ubound}
\end{align}
Let $\rho_{XM}$ be a c-q state with $X$ classical, such that $X=(X_1 \ldots X_k) \sim \theta$ and $\forall x\in \X:~  \rho_M^{x}= g(x)$. Consider,
\begin{align*}
\sum_i u_i(\theta) &= \sum_i \min_{\gamma_i} \expec{(T,x_T) \leftarrow \alpha, x_i \leftarrow\theta_i}{ \relent{g_{\theta}(x_i, x_T)}{g_{\theta_{-i},\gamma_i}(x_T)}}\\
    &\le \sum_i \expec{(T,x_T) \leftarrow \alpha, x_i \leftarrow \theta_i}{ \relent{g_{\theta}(x_i,x_T)}{g_{\theta_{-i},\theta_i}(x_T)}}\\
    &= \sum_i \expec{(T,x_T)\leftarrow\alpha}{\condmutinf{X_i}{M}{X_T=x_T}_\rho} &\mbox{(Fact~\ref{fact:inf})}\\
    &\le  \expec{(T,x_T)\leftarrow\alpha}{\condmutinf{X}{M}{X_T=x_T}_\rho} & \mbox{(Fact~\ref{fact:chaininf})}\\
    &\le \c(g) \enspace. & \mbox{(Definition~\ref{def:cap})} 
\end{align*}
Combining this with Eq.~\eqref{eq:ubound} concludes the desired. 
\end{proof}

\section{Chain-rule for the capacity of quantum channels}
\label{sec:chainquantum}
\subsection*{Capacity}
Let $\H_A, \H_B, \H_C$ be Hilbert spaces such that $A \equiv B$. Let $g: \L(B)\rightarrow \L(C)$ be a quantum channel. For $\rho \in \D(A)$, let $\hat{\rho}\in \D(AB)$ represent the canonical purification of $\rho$.
\begin{definition}[Capacity]
\label{def:capq}
The capacity of $g$ is defined as 
$$\c(g) \defeq \max_{\rho \in \D(A)} \mutinf{A}{C}_{g(\hat{\rho})} \enspace.$$
\end{definition}
We define the following notion of a capacity-dual.
\begin{definition}[Capacity-dual]
The capacity-dual of $g$ is defined as 
$$\tc(g) \defeq \min_{\gamma \in \D(A)}\max_{\rho \in \D(A)} \relent{g(\hat{\rho})_{AC}}{\rho_A \otimes g(\hat{\gamma})_C}\enspace.$$
\end{definition}
We show that capacity and capacity-dual are the same.
\begin{lemma} \label{fact:capdualq}
\begin{align*}
    \c(g) &= \max_{\rho \in \D(A)} \min_{\gamma \in \D(A)} \relent{g(\hat{\rho})_{AC}}{\rho_A \otimes g(\hat{\gamma})_C} \\
    &= \min_{\gamma \in \D(A)}\max_{\rho \in \D(A)} \relent{g(\hat{\rho})_{AC}}{\rho_A \otimes g(\hat{\gamma})_C} = \tc(g)\enspace.
\end{align*}
\end{lemma}
\begin{proof}
Let $\rho_A= p \rho^0_A+ (1-p) \rho^1_A$, for some $p\in[0,1]$. Let $\E$ be the map obtained form Fact~\ref{fact:PGM} and let $\sigma_{XAC}= \E(g(\hat{\rho}))_{AC}$. Note that $\sigma_{XA} = \E(\rho_A)$.  Consider
\begin{align}
    \relent{g(\hat{\rho})_{AC}}{\rho_A \otimes g(\hat{\gamma})_C} &\ge \relent{\sigma_{XAC}}{\sigma_{XA} \otimes g(\hat{\gamma})_C}  & \mbox{(Fact~\ref{fact:mono})} \nonumber\\
    &= p     \relent{g(\hat{\rho^0})_{AC}}{\rho^0_A \otimes g(\hat{\gamma})_C} + (1-p)     \relent{g(\hat{\rho^1})_{AC}}{\rho^1_A \otimes g(\hat{\gamma})_C}\enspace. &\mbox{(Fact~\ref{fact:relchain})} \label{eq:relconv} 
\end{align}
Consider,
\begin{align*}
    \c(g) &= \max_{\rho \in \D(A)} \mutinf{A}{C}_{g(\hat{\rho})} \\
    &= \max_{\rho \in \D(A)}  \relent{g(\hat{\rho})_{AC}}{\rho_A \otimes g(\hat{\rho})_C} &\mbox{(Fact~\ref{fact:inf})}\\
    & = \max_{\rho \in \D(A)}  \min_{\gamma \in \D(A)} \relent{g(\hat{\rho})_{AC}}{\rho_A \otimes g(\hat{\gamma})_C}&\mbox{(Fact~\ref{fact:inf})}\\
  &=\min_{\gamma \in \D(A)}\max_{\rho \in \D(A)} \relent{g(\hat{\rho})_{AC}}{\rho_A \otimes g(\hat{\gamma})_C} &\mbox{(Fact~\ref{fact:rel-conv}, Eq.~\eqref{eq:relconv}, Fact~\ref{fact:minmax})}\\
    &=\tc(g)\enspace.
\end{align*}
\end{proof}
\subsection*{Chain-rule}
Let $\H_A = \H_{A_1} \otimes \ldots \otimes \H_{A_k}, \H_B = \H_{B_1} \otimes \ldots \otimes \H_{B_k}$ and $\H_C$ be Hilbert spaces such that $\forall i\in [k]: A_i \equiv B_i$. Let $g: \L(B)\rightarrow \L(C)$ be a quantum channel.  Let $\A=\D(A_1) \times \ldots \times \D(A_k)$. We use $(\rho_1,  \ldots, \rho_k)  \in \A$ to represent $(\rho_1 \otimes \ldots \otimes  \rho_k)$. For $i\in[k]$, let $\hat{\rho_i}\in \D(A_iB_i)$ represent the canonical purification of $\rho_i$. For $i\in[k]$ and $\theta \in \D(A)$, define channel $g^i_{\theta}: \L(B_i) \rightarrow \L(C)$ given by $g^i_\theta(\rho_i)=g(\rho_i,\theta_{-i})$. We show the following.
\begin{theorem}[Chain-rule for quantum capacity] 
\begin{align*}
\c(g) ~\ge~   \min_{\theta \in \A}\sum_{i=1}^k \c(g^i_\theta)\enspace.
\end{align*}
\end{theorem}
\begin{proof}
For all $i\in[k] , \rho= (\rho_1, \ldots, \rho_k) \in \A$ , define 
\begin{displaymath}
    u_i(\rho) = \mutinf{A_i}{C}_{g(\hat{\rho})} \enspace. 
\end{displaymath}
For all $\rho, \rho'_i, \rho''_i, p\in [0,1]$,
\begin{align}
 &u_i(p\rho'_i + (1-p)\rho_i'', \rho_{-i}) \nonumber\\
 & =  \mutinf{A_i}{C}_{g(\widehat{p\rho'_i + (1-p)\rho_i''},\widehat{ \rho_{-i}})}  \nonumber \\
& \ge p \cdot \mutinf{A_i}{C}_{g(\hat{\rho'_i}, \hat{\rho_{-i}})} + (1-p) \cdot \mutinf{A_i}{C}_{g(\hat{\rho''_i}, \hat{\rho_{-i}})} &\mbox{(Fact~\ref{fact:PGM},  Fact~\ref{fact:mono}, Fact~\ref{fact:chaininf})} \nonumber\\
& = p \cdot u_i(\rho'_i, \rho_{-i}) + (1-p) \cdot u_i(\rho_i'',\rho_{-i}) \enspace. \label{eq:nashcond2}
\end{align}
From Eq.~\eqref{eq:nashcond2} and Fact~\ref{fact:nash} (by letting $\forall i: (\A_i, u_i) \leftarrow (\D(A_i), u_i)$), we get $\theta = (\theta_1, \ldots, \theta_k) \in \A$ such that, 
\begin{align*}
\forall i:~u_i(\theta) &= \max_{\rho_i\in \D(A_i)} u_i(\rho_i,\theta_{-i})  = \c(g^i_\theta) \enspace.
\end{align*}
Consider,
\begin{align*}
    \sum_i u_i(\theta) &= \sum_i \mutinf{A_i}{C}_\theta \\
    & \le \mutinf{A}{C}_\theta &\mbox{(Fact~\ref{fact:chaininf})} \\
    & \le \c(g) \enspace. & \mbox{(Definition~\ref{def:capq})} 
\end{align*}
This concludes the desired. 
\end{proof}

\subsection*{Acknowledgment}
This work is supported by the NRF RF Award No. NRF-NRFF2013-13; the Prime Minister's Office, Singapore and the Ministry of Education, Singapore, under the Research Centres of Excellence program and by Grant No. MOE2012-T3-1-009; the NRF2017-NRF-ANR004 {\em VanQuTe} Grant and the {\em VAJRA} Grant, Department of Science and Technology, Government of India.


\end{document}